\documentclass[%
 aip,
 amsmath,amssymb,
 reprint,%
]{revtex4-1}

\usepackage{graphicx}
\usepackage{dcolumn}
\usepackage{bm}

\usepackage[utf8]{inputenc}
\usepackage[T1]{fontenc}
\usepackage{mathptmx}
\usepackage{etoolbox}

\usepackage{float}
\usepackage{subfigure}
\usepackage[colorlinks, breaklinks, linkcolor=blue, anchorcolor=blue, citecolor=blue, urlcolor=blue]{hyperref}
\usepackage{dcolumn}
\usepackage{bm}

\usepackage[utf8]{inputenc}
\usepackage[T1]{fontenc}
\usepackage{mathptmx}
\usepackage{etoolbox}
\usepackage{tikz}
\usetikzlibrary{shapes}
\makeatletter
\def\@email#1#2{%
 \endgroup
 \patchcmd{\titleblock@produce}
  {\frontmatter@RRAPformat}
  {\frontmatter@RRAPformat{\produce@RRAP{*#1\href{mailto:#2}{#2}}}\frontmatter@RRAPformat}
  {}{}
}%
\makeatother
\begin{document}

\preprint{AIP/123-QED}
	
	\title{Doppler-enhanced superheterodyne Rydberg microwave receiver}
	\author{Yuwen Yin}    
	\author{Ruimin Chen}
        \author{Shibing Ji}    
        \author{Jinlian Hu}
        \author{Shaofeng Wang}
                   	\author{Yunhui He}
	\affiliation{State Key Laboratory of Quantum Optics Technologies and Devices, Institute of Laser Spectroscopy, Shanxi University, Taiyuan 030006, China}
         
	\author{Jingxu Bai}
    \affiliation{State Key Laboratory of Quantum Optics Technologies and Devices, Institute of Laser Spectroscopy, Shanxi University, Taiyuan 030006, China}
    \affiliation{Collaborative Innovation Center of Extreme Optics, Shanxi University, Taiyuan 030006, China}

\author{Xiao-Qiang Shao*}
\email[Author to whom correspondence should be addressed: xqshao@nenu.edu.cn]{}
\affiliation{
Center for Quantum Science and School of Physics, Northeast Normal University, Changchun, Jilin 130024, China}    
\affiliation{Institute of Quantum Science and Technology, Yanbian University, Yanji 133002, China}

	\author{Yuechun Jiao*}
	\email[Author to whom correspondence should be addressed: ycjiao@sxu.edu.cn]{}

	\author{Jianming Zhao*}
	\email[Author to whom correspondence should be addressed: zhaojm@sxu.edu.cn]{}
    
	\affiliation{State Key Laboratory of Quantum Optics Technologies and Devices, Institute of Laser Spectroscopy, Shanxi University, Taiyuan 030006, China}	\affiliation{Collaborative Innovation Center of Extreme Optics, Shanxi University, Taiyuan 030006, China}
	
	\date{\today}
	
\begin{abstract}
We report the enhanced sensitivity of the Rydberg microwave (MW) receiver by exploiting the Doppler effect in a vapor cell. A two-photon Rydberg ladder scheme is implemented via the co-propagation of probe and coupling lasers, which enhances the Doppler effect. When an MW field is applied, microwave dressing modifies the velocity-dependent resonance condition, enabling stronger contributions from atoms with non-zero velocities and leading to an enhancement of the EIT transmission. Based on this mechanism, we achieve a sensitivity of $35.1\ \mathrm{nV\ cm^{-1}\ Hz^{-1/2}}$ using the heterodyne technique, which is 1.5 times better than that obtained in the counter-propagating configuration. Meanwhile, the required local oscillator (LO) field is reduced by a factor of 17.6 compared with the counter-propagating configuration, which is advantageous for applications requiring minimal radiation and low power consumption. Moreover, the co-propagating configuration is more amenable to integration or portable sensing platforms because multiple laser fields can be delivered through a single optical fiber. 
        
\end{abstract}
	
	
	\begin{quotation}
	\end{quotation}
	
	\maketitle
	
Rydberg atoms have emerged as a promising platform for microwave (MW) field measurement~\cite{schlossberger2024a} due to their unique properties of large polarizability and MW transition dipole moment, and abundant Rydberg transitions.~\cite{gallagher1994rydberg,shao2024rydberg} An optical electromagnetically induced transparency (EIT)~\cite{Mohapatra2007} and Autler-Townes (AT) splitting~\cite{Tanasittikosol2011} spectroscopy have been employed to measure the electric fields with a broad frequency range from MHz to THz.~\cite{Fan2015,chen2022,li2023,lin2025} Developments include measurements of MW fields,~\cite{Sedlacek2012, Sedlacek13, SimonsMT2019, schmidt2025} imaging,~\cite{Holloway2014, Wade2017, downes2020b} and wireless communication.~\cite{meyer2018, deb2018b, Jiao2019, Song2019, Anderson2021, Holloway2019, Liu2022b} The sensitivity of the Rydberg receiver has been greatly improved to 55~nVcm$^{-1}$Hz$^{-1/2}$ using the Rydberg-atom superheterodyne technique.~\cite{Jing2020} While the sensitivity can be further improved through complementary methods, such as adding a repumping laser,~\cite{Prajapati2021} using the thermal resonance-enhanced transparency,~\cite{hu2025} placing the atom in a resonant MW cavity,~\cite{Holloway:2022qwb, zhou2025a} and designing an all-optical superheterodyne configurations,~\cite{borowka2025, zhang2026} sensitivity is ultimately limited in a large part by the EIT linewidth, which is mainly broadened by the Doppler effect. 

For alkali-atom-based Rydberg receivers with two-photon excitation, a counter-propagating configuration of the probe and coupling lasers is commonly used to reduce Doppler effects. Nevertheless, residual Doppler mismatch still limits the EIT linewidth to the megahertz level.~\cite{su2022a} To further mitigate thermal effects, a three-photon excitation scheme is employed to minimize the wave-vector mismatch among the interacting laser fields, thereby markedly suppressing residual Doppler broadening and enabling the EIT linewidth to reach the sub-megahertz regime.~\cite{bohaichuk2023c, prajapati2023} Another approach is to excite hot atoms with short and intense lasers,~\cite{baluktsian2013, bai2020} for which thermal motion becomes relatively insignificant on the nanosecond (or shorter) timescale.
Alternatively, narrow-linewidth Rydberg EIT can be realized in a cold atomic cloud.~\cite{liao2020a}

On the other hand, the Doppler effect may be regarded as a natural manifestation of motion that connects thermal energy with optical frequency through laser spectroscopy, thereby enabling the realization of non-reciprocal optical devices,~\cite{zhang2018a, liang2020, dong2021} atomic frequency combs,~\cite{aumiler2005, afzelius2009, main2021} and synchronization.~\cite{wadenpfuhl2023} In view of the foregoing, it is natural to raise the question of whether the Doppler effect in atomic vapor could find its applications in MW measurement.

In this work, we present the Doppler-enhanced sensitivity of the MW receiver in a vapor cell. We consider a Rydberg two-photon ladder scheme, where a probe and coupling laser co-propagate through the vapor cell to excite the ground state $|g\rangle$ to the Rydberg state $|r_1\rangle$ via an intermediate state $|e\rangle$. The co-propagation configuration enhances the Doppler effects.~\cite{su2025} When an MW field resonant with the transition between $|r_1\rangle$ and another Rydberg state $|r_2\rangle$ is applied, the microwave dressing modifies the velocity-dependent resonance condition, which increases the EIT transmission through the enhanced contribution of atoms with non-zero velocities over a certain range. 
Utilizing this effect, we achieve a sensitivity of $35.1\ \mathrm{nV\ cm^{-1}\ Hz^{-1/2}}$ using the heterodyne technique, which is 1.5 times better than that of the counter-propagating configuration. Meanwhile, the strength of the local oscillator (LO) field is $0.36\ \mathrm{mV\ cm^{-1}}$, which is 17.6 times lower than that of the counter-propagating configuration. The low LO field 
is significant for applications where radiation must be minimized and power consumption is a critical constraint. Additionally, compared with the counter-propagating configuration, the co-propagating configuration offers greater potential for integrated implementation, as different laser beams can be combined and delivered through a single optical fiber.


	\begin{figure}[htbp]
		\vspace{-1ex}
		\centering		\includegraphics[width=1\linewidth]{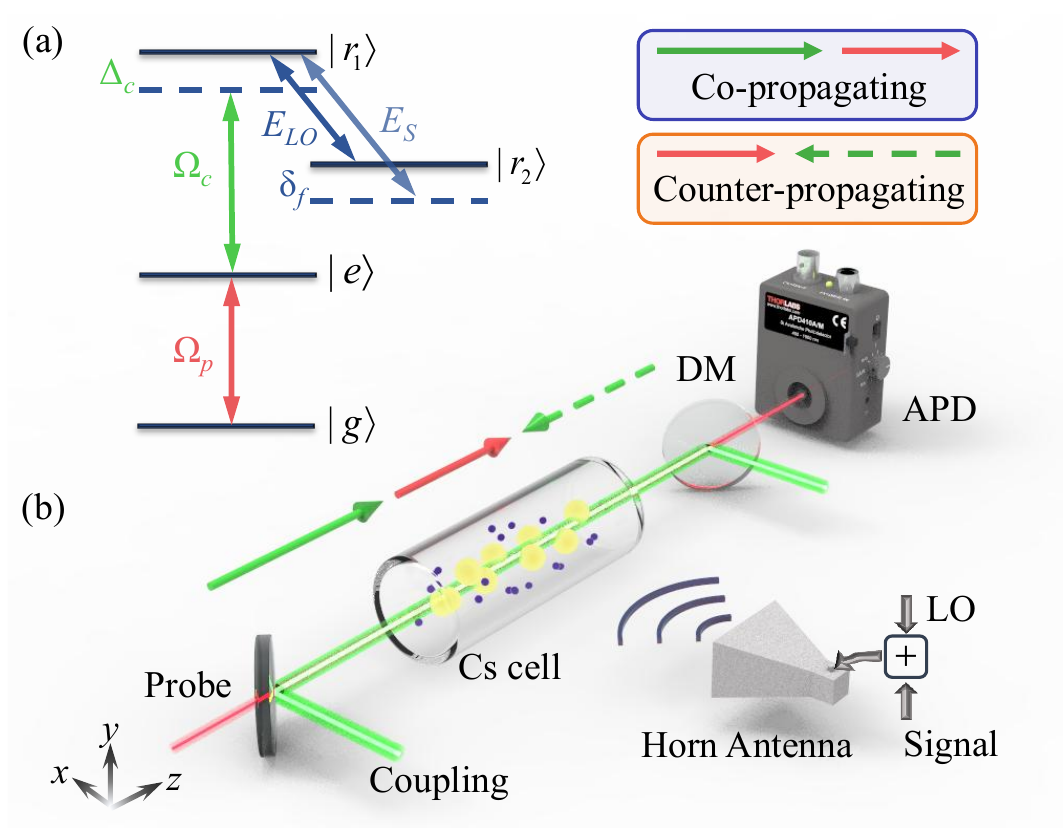}
		\vspace{-1ex}
		\caption{(a) Energy-level diagram of Rydberg EIT-AT.
        (b) Sketch of the experimental setup. An 852~nm probe laser (Rabi frequency $\Omega_p$) and a 509~nm coupling laser ($\Omega_c$) drive the ground atoms $|g\rangle$ to the Rydberg state $|r_1\rangle$ via an intermediate state $|e\rangle$. A LO field $E_{LO}$ drives the resonant transition of $|r_{1}\rangle\to |r_{2}\rangle$, while the signal field $E_S$ has a frequency difference $\delta_f$ with the LO field. The coupling laser 
        co-propagates (solid green arrow) or counter-propagates (dashed green arrow) with the probe laser. The transmission of the probe laser is detected by an avalanche photodiode (APD) after a dichroic mirror (DM). The LO field and the signal field are transmitted to the vapor cell by a horn antenna.      
        }
		\label{Fig.1}
	\end{figure}

The experimental setup and the relevant $^{133}$Cs Rydberg EIT-AT four-level diagram are illustrated in Fig.~\ref{Fig.1}. 
The experiments are performed in a $\phi$ 25~mm $\times$ 75~mm cylindrical cesium room-temperature vapor cell, where a probe laser (852~nm) and a coupling laser (509~nm) co-propagate and overlap through the cell. The probe laser resonantly couples the transition between the ground state $|g\rangle = |6S_{1/2}, F = 4\rangle$ and the excited state $|e\rangle = |6P_{3/2}, F^\prime = 5\rangle$ with an effective Rabi frequency $\Omega_p\approx2\pi\times 10.87\ \mathrm{MHz}$, estimated by taking into account the transverse Gaussian profile of the probe beam. The coupling laser further excites the atoms to the Rydberg state $|r_1\rangle = |48D_{5/2}\rangle$ with an effective  Rabi frequency $\Omega_c \approx 2\pi \times 4.47\ \mathrm{MHz}$. 
The transmission of the probe laser is detected by an APD. The $1/e^2$ beam waists of the probe and coupling lasers are 120~$\mu$m and 160~$\mu$m, respectively. Both laser polarizations are along the y-axis.
Two MW fields, denoted as a LO field $E_{LO}$ and a weak signal field $E_{SIG}$, are simultaneously emitted from a horn antenna (A-info LB-20180SF) incident on the cell with the same polarization as the lasers and propagating perpendicular to the lasers. The LO field frequency is resonant with the transition of $|48D_{5/2}\rangle\to |r_2\rangle = |49P_{3/2}\rangle$ at a frequency of 6.497~GHz and the transition dipole moment $\mu$ of 1509.07~$ea_{0}$, while the signal field has a $\delta_f$ = 30~kHz detuning. When the two MW fields interact simultaneously with the Rydberg atoms, atomic mixing generates a 30~kHz interference beat frequency in the transmission spectrum, which is detected by a spectrum analyzer (Rohde \& Schwarz FSVA3013). For comparison, we also perform experiments with the probe and coupling lasers in a counter-propagation configuration, in which we keep all other experimental parameters constant. This configuration is a common arrangement in most experiments, which achieves the maximum suppression of Doppler effects.~\cite{su2025}


\begin{figure*}[htbp]
		\vspace{-1ex}
		\centering		\includegraphics[width=0.95\textwidth]{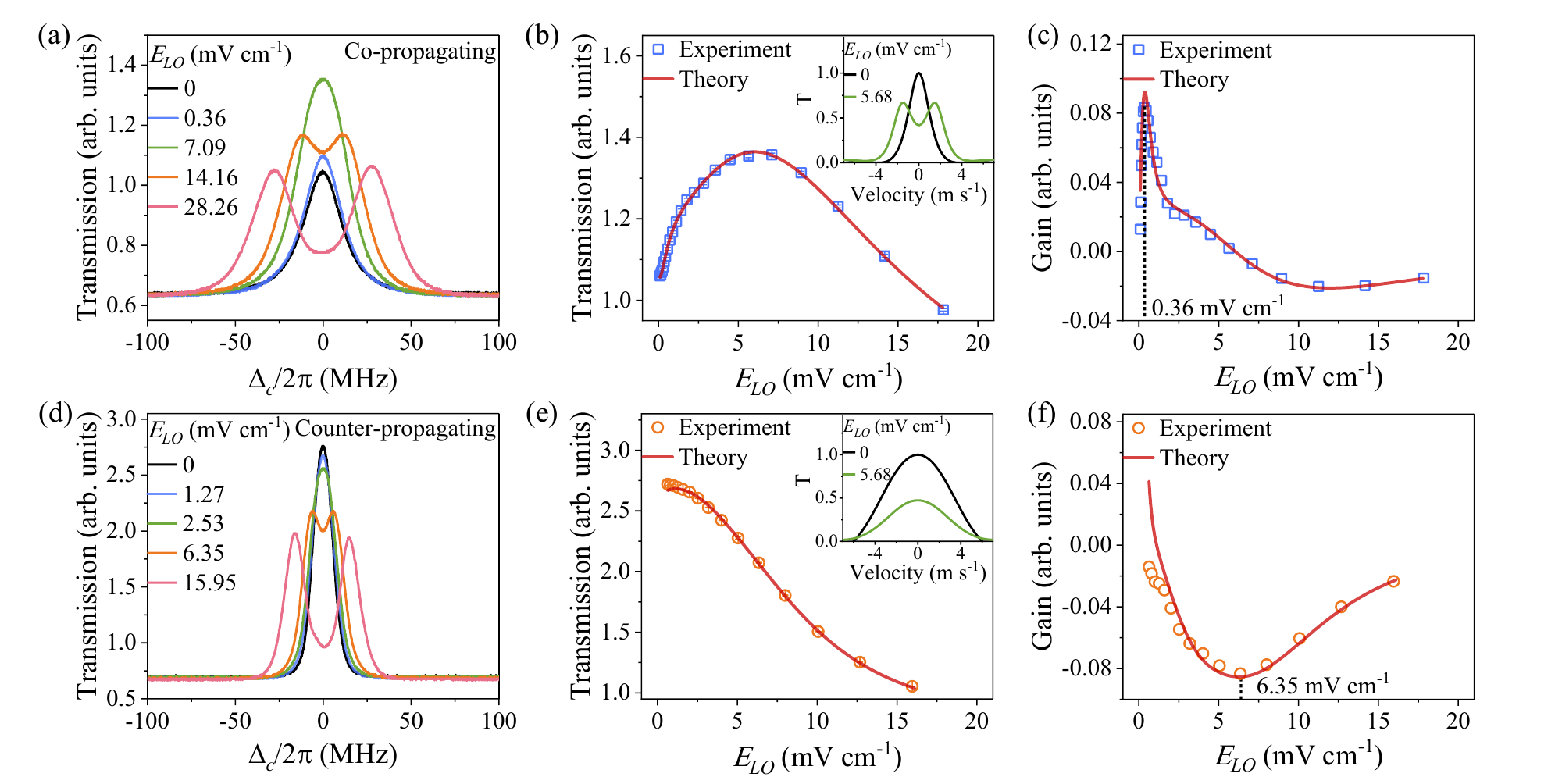}
		\vspace{-1ex}
		\caption{(a) Experimental scanned Rydberg EIT spectra at indicated strengths of $E_{LO}$ fields in co-propagation configuration. (b) and (c) The dependence of the EIT amplitude and its derivative on the MW field strength at $\Delta_c = 0$. 
         The inset in (b) demonstrates the transmission rate (T) of the different velocity classes of atoms at $E_{LO} =$ 0 and 5.68$\ \mathrm{mV\ cm^{-1}}$. (d-f) Scanned Rydberg EIT spectra, EIT amplitude, and its derivative as a function of the MW field strength, similar to (a-c), except for the counter-propagating configuration. 
        The vertical dashed lines mark the maximum gain point. The solid lines in (b) and (e) show our calculations with parameters $\Omega_p = 2\pi \times 6.62 \ (9.89) \ \mathrm{MHz}$, $\Omega_c = 2\pi \times 5.69 \ (4.32)\ \mathrm{MHz}$, Rydberg state dephasing rate $\Gamma_r = 2\pi \times 15 \ (20) \,\mathrm{kHz}$, and transit rate $\Gamma_{\rm transit} = 2\pi \times 300 \ (150)\,\mathrm{kHz}$. 
        }
		\label{Fig.2}
	\end{figure*}

With the probe and coupling beams co-propagating, we demonstrate the scanning Rydberg EIT-AT spectra as a function of coupling laser detuning $\Delta_c$ for various strengths of $E_{LO}$ fields, as shown in Fig.~\ref{Fig.2}(a). The black curve shows the three-level EIT spectrum without the $E_{LO}$ field. When a small MW field of $E_{LO} = 0.36\ \mathrm{mV\ cm^{-1}}$ is applied, we observe an increase in the EIT amplitude accompanied by a broadening of the EIT linewidth. As the MW field is increased to $E_{LO} = 7.09\ \mathrm{mV\ cm^{-1}}$, we surprisingly observe that the EIT amplitude increases by a factor of 1.76 compared with the MW-free case, which is markedly different from the behavior previously reported in counter-propagating configurations. With a further increase in the MW field strength, the EIT transmission begins to decrease and eventually develops two split peaks. To investigate the dependence of the EIT amplitude on the MW field strength, we perform a series of such measurements over an \textit{E}-field range of 0 $\sim$ 18.0 $\mathrm{mV\ cm^{-1}}$ and extract the amplitude of the spectra as a function of $E_{LO}$ at $\Delta_c =$ 0. The results are demonstrated in Fig.~\ref{Fig.2}(b), which shows that the spectrum peak begins to increase until it reaches a maximum at $E_{LO} = 7.09\ \mathrm{mV\ cm^{-1}}$ and then decreases. 
Subsequently, we take the derivative of the transmission amplitude versus the $E_{LO}$ field, which depicts its MW-optical gain, as shown in Fig.~\ref{Fig.2}(c). The maximum gain (the largest slope) 
occurs at $E_{LO} = 0.36\ \mathrm{mV\ cm^{-1}}$. 

For comparison, we demonstrate the influence of the MW field on the EIT spectra in the counter-propagating configuration of the probe and coupling beams, as shown in Fig.~\ref{Fig.2}(d). At a small MW electric field, such as $E_{LO} = 1.27\ \mathrm{mV\ cm^{-1}}$ and $2.53\ \mathrm{mV\ cm^{-1}}$, the MW electric fields cause broadening and reduction of the EIT peak. At an intermediate MW electric field $E_{LO} = 6.35\ \mathrm{mV\ cm^{-1}}$, the EIT spectrum splits into two peaks with the appearance of an absorption dip at $\Delta_c =$ 0. By further increasing the MW electric field, the separation between the two EIT peaks is proportional to the MW electric field strengths.~\cite{Sedlacek2012} Similarly, we extract the amplitude of the spectra and its derivative as a function of $E_{LO}$ in a range of 0 $\sim$ 16.0 $\mathrm{mV\ cm^{-1}}$, as shown in Fig.~\ref{Fig.2}(e) and (f). 
Especially, the curve has the largest slope at the value of $E_{LO} =$ 6.35~$\mathrm{mV\ cm^{-1}}$. Comparing the electric field values corresponding to the maximum slope in the co- and counter-propagating configurations, the value in the co-propagating case is 17.6 times smaller than that in the counter-propagating configuration.

To model our results, we start from the full four-level ladder system. In the rotating-wave approximation, the system is described by the Hamiltonian
\begin{eqnarray}
H/\hbar &= &-\Delta_p |e\rangle\langle e| - \Delta_{c} |r_1\rangle\langle r_1| - \Delta_{LO} |r_2\rangle\langle r_2|\\&&
+ (\frac{\Omega_p}{2}|g\rangle\langle e|\nonumber
+ \frac{\Omega_c}{2}|e\rangle\langle r_1|
+ \frac{\Omega_{LO}}{2}|r_1\rangle\langle r_2|
+ \mathrm{H.c.}),    
\end{eqnarray}
where $\hbar$ is the reduced Planck constant, $\Omega_p$, $\Omega_c$, and $\Omega_{LO}$ denote the probe, coupling, and MW Rabi frequencies, respectively. $\Delta_p$, $\Delta_{c}$, and $\Delta_{LO}$ are the detunings between the frequencies of the driving fields and atomic transitions. The open-system dynamics are governed by the Lindblad master equation
\begin{equation}\label{master}
    \dot{\rho}=-\frac{i}{\hbar}[H,\rho]+\mathcal{L}_{\rm sp}[\rho]+\mathcal{L}_{\rm deph}[\rho]+\Gamma_{\rm transit}(\rho_{\rm g}-\rho),
\end{equation}
where $\rho$ is the density matrix and $\rho_{\rm g}$ denotes the ground-state population matrix. The three dissipative terms describe spontaneous emission, pure dephasing, and transit-time relaxation, respectively. The transit-time relaxation arises from the finite interaction duration, as Rydberg atoms move in and out of the interaction volume. It is given by  $\Gamma_{\rm transit}$ = $v/(2w_0)$, where $v$ is the average velocity of Cs atoms and $w_0$ is the Gaussian beam waist. In a thermal vapor, each atomic velocity class experiences different Doppler shifts. By numerically solving the Lindblad master equation~(\ref{master}), we can obtain the EIT transmission amplitude at the resonant point as a function of $E_{LO}=\hbar\Omega_{LO}/\mu$, as shown by the solid lines in Fig.~\ref{Fig.2}(b) and~\ref{Fig.2}(e), which are in good agreement with the experimental data.

To understand the mechanism of enhanced transmission under the MW field, we develop a dressed state model, where the MW-induced hybridization of the two Rydberg states into a pair of dressed modes,
with eigenfrequencies separated by the MW Rabi frequency $\Omega_{LO}$. 
These dressed modes define two excitation pathways in the EIT process.
When both the probe and coupling fields are locked on resonance in the laboratory frame, i.e., $\Delta_p=\Delta_c=0$, atoms with velocity $v$ experience a Doppler-shifted effective Rydberg detuning,
\begin{equation}
\Delta_r(v)=-(k_p \pm k_c)v \pm \frac{\Omega_{LO}}{2},
\end{equation}
where the first $\pm$ sign corresponds to the propagation configuration, whereas the second $\pm$ sign denotes the eigenenergy shifts of the two MW-dressed states. 
This expression shows that the Doppler shift can partially compensate for the MW-induced splitting for one dressed mode while further detuning the other.

As a result, for a given velocity class, one dressed mode can be brought close to resonance and dominate the optical response, while the other becomes effectively inactive. This Doppler-assisted mode selection governs the observed spectral behavior.
In the co-propagating configuration, the Doppler shifts add constructively, yielding a large velocity-dependent detuning $(k_p+k_c)v$. This enables efficient compensation of the MW splitting, enhancing the EIT interference at moderate $\Omega_{LO}$. As illustrated in the inset of Fig.~\ref{Fig.2}(b), although the contribution from the zero-velocity class is reduced due to MW dressing, the transmission from atoms with finite velocities is significantly enhanced. As a result, the overall integrated contribution over the velocity distribution increases, leading to an enhanced EIT signal. At larger MW fields, the splitting exceeds the Doppler compensation range, leading to reduced transparency and a non-monotonic dependence on $\Omega_{LO}$. In contrast, in the counter-propagating configuration, the Doppler shifts largely cancel, resulting in a much weaker velocity dependence $(k_p-k_c)v$. The Doppler effect is then insufficient to compensate for the MW-induced splitting, so that both dressed modes remain off-resonant. As shown in the inset of Fig.~\ref{Fig.2}(e), the MW dressing mainly suppresses the central (near-zero velocity) contribution without providing sufficient enhancement from other velocity classes, resulting in an overall reduction of the integrated transmission. Consequently, the transmission exhibits only a slight initial increase at very small $\Omega_{LO}$, followed by a continuous decrease as the MW-induced splitting dominates. The phenomenon of the counter-propagating configuration has been investigated in detail.~\cite{Sedlacek2012} 

\begin{figure}[htbp]
		\centering
		\vspace{-1ex}
		\includegraphics[width=1\linewidth]{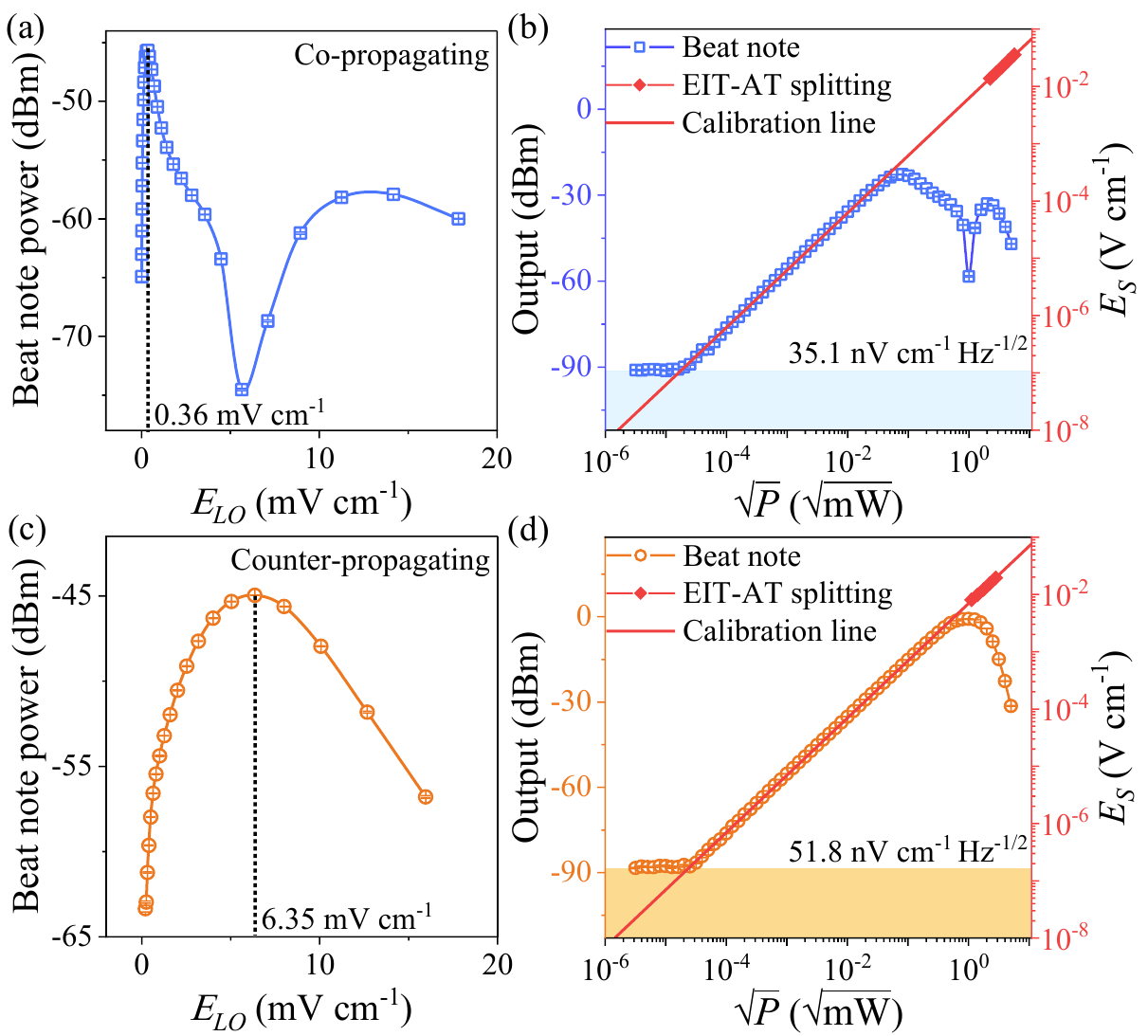}
		\vspace{-1ex}
		\caption{(a) and (c) The response of the system to the LO field with fixed signal field using superheterodyne measurement for the co-propagating and counter-propagating configurations, respectively. The vertical dashed lines mark the maximum output point. 
        (b) and (d) Measured sensitivity for the co-propagating (blue squares) and counter-propagating (yellow circles) configurations, respectively. The red diamonds show the EIT-AT splittings in a strong field region, and the red solid line shows the calibrated electric field. The blue and yellow shadow regions represent the noise level of the spectrum analyzer.
        }
		\label{Fig.3}
	\end{figure}

The enhanced transmission under MW field can be utilized to measure the MW field. 
At the point of maximum slope, the atoms exhibit maximum sensitivity to MW field variations, which we subsequently verify using the superheterodyne technique. By locking the coupling laser to the resonant point, i.e., $\Delta_c = 0$, we measure the response of the system to the LO field with the signal field fixed to $E_{S} = 19.59\ \mathrm{\mu V\ cm^{-1}}$. In the presence of both the LO field and the signal field, the transmission of the probe exhibits the beat note signal. We measure the beat note power as a function of the LO field $E_{LO}$ using the spectrum analyzer. In Fig.~\ref{Fig.3}(a) and (c), we present the output power of the spectrum analyzer for the co-propagating and counter-propagating configurations, respectively. We can see that the LO field has an optimal operating point, which is $E_{LO} = 0.36\ \mathrm{mV\ cm^{-1}}$ and $6.35\ \mathrm{mV\ cm^{-1}}$. The value is consistent with the value obtained from the slope measurement.

Then, we fix the strength of the LO field operating at the optimal point to measure the response of the system to the signal field using atomic superheterodyne. In Fig.~\ref{Fig.3}(b) and (d), we demonstrate the sensitivity measurements of the Rydberg receiver with a measurement time of 0.1~s. The results show that the system linearly responds to the signal fields with a linear dynamical range of 68~dB [blue squares in (b)] and 86~dB [yellow circles in (d)] for the co-propagating and counter-propagating configurations, respectively. The minimum detectable signal field is $111\ \mathrm{nV\ cm^{-1}}$ for the co-propagating configurations and $164\ \mathrm{nV\ cm^{-1}}$ for the counter-propagating configurations, which are determined when the power of the oscillation signal reaches the noise level of the spectrum analyzer (blue and yellow shadow regions). The corresponding sensitivities are $35.1\ \mathrm{nV\ cm^{-1}\ Hz^{-1/2}}$ and $51.8\ \mathrm{nV\ cm^{-1}\ Hz^{-1/2}}$, showing an enhancement of a factor of 1.5 using the co-propagating configurations. 
We find that, compared to the counter-propagating configuration, the co-propagating configuration exhibits higher field sensitivity but a less linear dynamical response range, due to a smaller linear response to the LO field.
The EIT-AT splitting in a strong field region (red diamonds) and the far-field formula $E_{\mathrm{FF}} = {F\sqrt{30P\cdot g}}/{d}$ (red solid line) are used to determine the strength of the signal fields at its corresponding setting power of $\sqrt{P}$, where $g$ is a gain factor of the antenna, $d$ is the distance between the horn antenna port and the center of the cesium cell, and $F$ is the cell perturbation factor. 

In summary, we demonstrate a Doppler-enhanced Rydberg MW receiver in a thermal vapor cell using a co-propagating probe-coupling configuration. An applied MW field modifies the velocity-dependent resonance condition, which results in an increase in the EIT transmission as atoms with non-zero velocities participate more strongly in the EIT process. This effect leads to an enhanced response to weak MW fields and enables sensitive MW detection. Operating at the optimal LO field point, we verify this behavior using the atomic superheterodyne technique and obtain a sensitivity of $35.1\ \mathrm{nV\ cm^{-1}\ Hz^{-1/2}}$, about 1.5 times better than that in the counter-propagating configuration. Meanwhile, the optimal local oscillator field is reduced by a factor of 17.6, which is especially beneficial, for example, in integrated or portable sensing platforms. In addition, the co-propagating configuration also facilitates system integration since multiple laser fields can be delivered through a single optical fiber. These results provide a promising route toward compact and low-power atomic MW sensing systems.

	\vspace{5mm}
	This work is supported by the National Natural Science Foundation of China (No. 12241408, U2341211, 12120101004, and 12504304); Quantum Science and Technology-National Science and Technology Major Project (No. 2023ZD0300902); and Fundamental Research Program of Shanxi Province (No. 202303021224007, 202503021212074, 202303021212002). X. Q. Shao was supported by the National Natural Science Foundation (Grant No. 12174048).


\vspace{12pt}
\section*{AUTHOR DECLARATIONS}
\subsection*{Conflict of Interest}
The authors have no conflicts to disclose.

\vspace{12pt}
\subsection*{Author Contributions}
	
	\section*{Data Availability}
	The data that support the findings of this study are available from the corresponding author upon reasonable request.
	

	\bibliography{main}

@article{zhang2018a,
  title = {Thermal-Motion-Induced Non-Reciprocal Quantum Optical System},
  author = {Zhang, Shicheng and Hu, Yiqi and Lin, Gongwei and Niu, Yueping and Xia, Keyu and Gong, Jiangbin and Gong, Shangqing},
  year = {2018},
  journal = {Nat. Photonics},
  volume = {12},
  number = {12},
  pages = {744--748},
  issn = {1749-4885, 1749-4893},
  doi = {10.1038/s41566-018-0269-2},
  urldate = {2024-03-27},
  langid = {english}
}

@article{shao2024rydberg,
    title = {{R}ydberg superatoms: An artificial quantum system for quantum information processing and quantum optics},
    author = {Shao, Xiao-Qiang and Su, Shi-Lei and Li, Lin and Nath, Rejish and Wu, Jin-Hui and Li, Weibin},
    journal = {Appl. Phys. Rev.},
    volume = {11},
    number = {3},
    pages = {031320},
    year = {2024},
    month = {08},
    doi = {10.1063/5.0211071},
    url = {https://doi.org/10.1063/5.0211071}
}

@article{liang2020,
  title = {Collision-{{Induced Broadband Optical Nonreciprocity}}},
  author = {Liang, Chao and Liu, Bei and Xu, An-Ning and Wen, Xin and Lu, Cuicui and Xia, Keyu and Tey, Meng Khoon and Liu, Yong-Chun and You, Li},
  year = {2020},
  journal = {Phys. Rev. Lett.},
  volume = {125},
  number = {12},
  pages = {123901},
  issn = {0031-9007, 1079-7114},
  doi = {10.1103/PhysRevLett.125.123901},
  urldate = {2024-03-28},
  langid = {english}
}

@article{dong2021,
  title = {All-Optical Reversible Single-Photon Isolation at Room Temperature},
  author = {Dong, Ming-Xin and Xia, Ke-Yu and Zhang, Wei-Hang and Yu, Yi-Chen and Ye, Ying-Hao and Li, En-Ze and Zeng, Lei and Ding, Dong-Sheng and Shi, Bao-Sen and Guo, Guang-Can and Nori, Franco},
  year = {2021},
  journal = {Sci. Adv.},
  volume = {7},
  number = {12},
  pages = {eabe8924},
  issn = {2375-2548},
  doi = {10.1126/sciadv.abe8924},
  urldate = {2024-03-28},
  langid = {english}
}

@article{main2021,
  title = {Room Temperature Atomic Frequency Comb Storage for Light},
  author = {Main, D and Hird, T M and Gao, S and Walmsley, I A and Ledingham, P M},
  number = {12},
  pages = {2960--2963},
  volume = {46},
  month = {Jun},
  year = {2021},
  journal = {Opt. Lett.},
  url = {https://opg.optica.org/ol/abstract.cfm?URI=ol-46-12-2960},
  doi = {10.1364/OL.426753},
  langid = {english}
}

@article{aumiler2005,
  title = {Velocity Selective Optical Pumping of Rb Hyperfine Lines Induced by a Train of Femtosecond Pulses},
  author = {Aumiler, D. and Ban, T. and Skenderovi\ifmmode \acute{c}\else \'{c}\fi{}, H. and Pichler, G.},
  year = {2005},
  journal = {Phys. Rev. Lett.},
  langid = {english},
  volume = {95},
  issue = {23},
  pages = {233001},
  number = {4},
  month = {Nov},
  publisher = {American Physical Society},
  doi = {10.1103/PhysRevLett.95.233001},
  url = {https://link.aps.org/doi/10.1103/PhysRevLett.95.233001}
}

@article{afzelius2009,
  title = {Multimode Quantum Memory Based on Atomic Frequency Combs},
  author = {Afzelius, Mikael and Simon, Christoph and De Riedmatten, Hugues and Gisin, Nicolas},
  year = {2009},
  journal = {Phys. Rev. A},
  volume = {79},
  number = {5},
  pages = {052329},
  issn = {1050-2947, 1094-1622},
  doi = {10.1103/PhysRevA.79.052329},
  urldate = {2024-11-14},
  copyright = {http://link.aps.org/licenses/aps-default-license},
  langid = {english}
}

@article{prajapati2023,
  title = {Sensitivity Comparison of Two-Photon vs Three-Photon {{Rydberg}} Electrometry},
  author = {Prajapati, Nikunjkumar and Bhusal, Narayan and Rotunno, Andrew P. and Berweger, Samuel and Simons, Matthew T. and {Artusio-Glimpse}, Alexandra B. and Ju Wang, Ying and Bottomley, Eric and Fan, Haoquan and Holloway, Christopher L.},
  year = {2023},
  journal = {J. Appl. Phys.},
  volume = {134},
  number = {2},
  pages = {023101},
  issn = {0021-8979, 1089-7550},
  doi = {10.1063/5.0147827},
  urldate = {2024-01-20},
  langid = {english}
}

@article{Fan2015,
	doi = {10.1088/0953-4075/48/20/202001},
	url = {https://doi.org/10.1088/0953-4075/48/20/202001},
	year = 2015,
	month = {sep},
	publisher = {{IOP} Publishing},
	volume = {48},
	number = {20},
	pages = {202001},
	author = {Haoquan Fan and Santosh Kumar and Jonathon Sedlacek and Harald Kübler and Shaya Karimkashi and James P Shaffer},
	title = {Atom based {RF} electric field sensing},
	journal = {J. Phys. B: At. Mol. Opt. Phys.},
}

@ARTICLE{Wade2017,  
author={C. G. Wade and N. Šibalić and N.R. de Melo and J.M. Kondo and C.S.Adams and K. J. Weatherill},  journal={Nat. Photonics},   title={Real-time near-field terahertz imaging with atomic optical fluorescence},   year={2017},  volume={11},  issue={1},  pages={40-43},  doi={10.1038/nphoton.2016.214}}

@ARTICLE{Sedlacek2012,  
author={Jonathon A. Sedlacek  and Arne Schwettmann and Harald Kübler and Robert Löw and Tilman Pfau and James P. Shaffer },  journal={Nat. Phys.}, title={Microwave electrometry with Rydberg atoms in a vapour cell using bright atomic resonances},   year={2012},  volume={8},  number={12},  pages={819-824},  doi={10.1038/nphys2423}}

@article{Sedlacek13,
  title = {Atom-Based Vector Microwave Electrometry Using Rubidium Rydberg Atoms in a Vapor Cell},
  author = {J. A. Sedlacek and A. Schwettmann  and H. K\"ubler and J. P. Shaffer },
  journal = {Phys. Rev. Lett.},
  volume = {111},
  issue = {6},
  pages = {063001},
  numpages = {5},
  year = {2013},
  month = {Aug},
  publisher = {American Physical Society},
  doi = {10.1103/PhysRevLett.111.063001},
  url = {https://link.aps.org/doi/10.1103/PhysRevLett.111.063001}
}

@article{SimonsMT2019,
	author = {Simons, Matthew T. and Haddab, Abdulaziz H. and Gordon, Joshua A. and Holloway, Christopher L.},
	title = {A Rydberg atom-based mixer: Measuring the phase of a radio frequency wave},
	journal = {Appl. Phys. Lett.},
	volume = {114},
	number = {11},
	pages = {114101},
	DOI = {10.1063/1.5088821},
	year = {2019}
}

@article{Holloway2014,
author = {Holloway,Christopher L.  and Gordon,Joshua A.  and Schwarzkopf,Andrew  and Anderson,David A.  and Miller,Stephanie A.  and Thaicharoen,Nithiwadee  and Raithel,Georg },
title = {Sub-wavelength imaging and field mapping via electromagnetically induced transparency and Autler-Townes splitting in Rydberg atoms},
journal = {Appl. Phys. Lett.},
volume = {104},
number = {24},
pages = {244102},
year = {2014},
doi = {10.1063/1.4883635},
URL = { https://doi.org/10.1063/1.4883635}
}

@article{Jiao2019,
	author = {Jiao, Yuechun and Han, Xiaoxuan and Fan, Jiabei and Raithel, Georg and Zhao, Jianming and Jia, Suotang},
	title = {Atom-based receiver for amplitude-modulated baseband signals in high-frequency radio communication},
	journal = {Appl. Phys. Express},
	volume = {12},
	number = {12},
	pages = {126002},
	DOI = {10.7567/1882-0786/ab5463},
	year = {2019}
}

@article{Song2019,
	author = {Song, Z. and Liu, H. and Liu, X. and Zhang, W. and Zou, H. and Zhang, J. and Qu, J.},
	title = {Rydberg-atom-based digital communication using a continuously tunable radio-frequency carrier},
	journal = {Opt. Express},
	volume = {27},
	number = {6},
	pages = {8848-8857},
	DOI = {10.1364/OE.27.008848},
	year = {2019}
}

@article{Anderson2021,
	author = {Anderson, D. A. and Sapiro, R. E. and Raithel, G.},
	title = {An Atomic Receiver for AM and FM Radio Communication},
	journal = {IEEE Trans. Antennas Propag.},
	volume = {69},
	number = {5},
	pages = {2455-2462},
	DOI = {10.1109/TAP.2020.2987112},
	year = {2021}
}

@article{lin2025,
  title = {Terahertz Receiver Based on Room-Temperature {{Rydberg-atoms}}},
  author = {Lin, Ya-Yi and She, Zhen-Yue and Chen, Zhi-Wen and Li, Xian-Zhe and Zhang, Cai-Xia and Liao, Kai-Yu and Zhang, Xin-Ding and Chen, Jie-Hua and Huang, Wei and Yan, Hui and Zhu, Shi-Liang},
  year = {2025},
  journal = {Fundam. Res.},
  volume = {5},
  number = {3},
  pages = {970--974},
  issn = {26673258},
  doi = {10.1016/j.fmre.2023.02.019},
  urldate = {2025-09-29},
  langid = {english}
}

@book{gallagher1994rydberg,
  title = {Rydberg {{Atoms}}},
  author = {Gallagher, Thomas F.},
  year = {1994},
  series = {Cambridge {{Monographs}} on {{Atomic}}, {{Molecular}} and {{Chemical Physics}}},
  publisher = {{Cambridge University Press}},
  location = {{Cambridge}},
  doi = {10.1017/CBO9780511524530},
  url = {https://www.cambridge.org/core/books/rydberg-atoms/B610BDE54694936F496F59F326C1A81B},
  isbn = {978-0-521-02166-1}
}

@article{Holloway2019,
	author = {Holloway, C. L. and Simons, M. T. and Gordon, J. A. and Novotny, D.},
	title = {Detecting and Receiving Phase-Modulated Signals With a Rydberg Atom-Based Receiver},
	journal = {IEEE Antennas Wireless Propag. Lett.},
	volume = {18},
	number = {9},
	pages = {1853-1857},
	DOI = {10.1109/LAWP.2019.2931450},
	year = {2019},
}

@article{Liu2022b,
author = {Liu, Zong-Kai and Zhang, Li-Hua and Liu, Bang and Zhang, Zheng-Yuan and Guo, Guang-Can and Ding, Dong-Sheng and Shi, Bao-Sen},
doi = {10.1038/s41467-022-29686-7},
issn = {2041-1723},
journal = {Nat. Commun.},
month = {dec},
number = {1},
pages = {1997},
title = {{Deep learning enhanced Rydberg multifrequency microwave recognition}},
url = {http://arxiv.org/abs/2202.13617%0Ahttp://dx.doi.org/10.1038/s41467-022-29686-7 https://www.nature.com/articles/s41467-022-29686-7},
volume = {13},
year = {2022}
}

@article{schlossberger2024a,
  title = {Rydberg States of Alkali Atoms in Atomic Vapour as {{SI-traceable}} Field Probes and Communications Receivers},
  author = {Schlossberger, Noah and Prajapati, Nikunjkumar and Berweger, Samuel and Rotunno, Andrew P. and Artusio-Glimpse, Alexandra B. and Simons, Matthew T. and Sheikh, Abrar A. and Norrgard, Eric B. and Eckel, Stephen P. and Holloway, Christopher L.},
  journal = {Nat. Rev. Phys.},
  volume = {6},
  number = {10},
  pages = {606--620},
  issn = {2522-5820},
  doi = {10.1038/s42254-024-00756-7},
  url = {https://doi.org/10.1038/s42254-024-00756-7},
  year = {2024},
  langid = {english}
}

@article{bohaichuk2023c,
  title = {Three-Photon {{Rydberg-atom-based}} Radio-Frequency Sensing Scheme with Narrow Linewidth},
  author = {Bohaichuk, Stephanie M. and Ripka, Fabian and Venu, Vijin and Christaller, Florian and Liu, Chang and Schmidt, Matthias and K{\"u}bler, Harald and Shaffer, James P.},
  year = 2023,
  journal = {Phys. Rev. Appl.},
  volume = {20},
  number = {6},
  pages = {L061004},
  issn = {2331-7019},
  doi = {10.1103/PhysRevApplied.20.L061004},
  urldate = {2024-08-24},
  langid = {english}
}

@article{wadenpfuhl2023,
  title = {Emergence of {{Synchronization}} in a {{Driven-Dissipative Hot Rydberg Vapor}}},
  author = {Wadenpfuhl, Karen and Adams, C. Stuart},
  year = 2023,
  journal = {Phys. Rev. Lett.},
  volume = {131},
  number = {14},
  pages = {143002},
  issn = {0031-9007, 1079-7114},
  doi = {10.1103/PhysRevLett.131.143002},
  urldate = {2023-10-14},
  langid = {english}
}

@article{liao2020a,
  title = {Microwave Electrometry via Electromagnetically Induced Absorption in Cold {{Rydberg}} Atoms},
  author = {Liao, Kai-Yu and Tu, Hai-Tao and Yang, Shu-Zhe and Chen, Chang-Jun and Liu, Xiao-Hong and Liang, Jie and Zhang, Xin-Ding and Yan, Hui and Zhu, Shi-Liang},
  year = 2020,
  journal = {Phys. Rev. A},
  volume = {101},
  number = {5},
  pages = {053432},
  issn = {2469-9926, 2469-9934},
  doi = {10.1103/PhysRevA.101.053432},
  urldate = {2025-09-04},
  langid = {english}
}

@article{baluktsian2013,
  title = {Evidence for {{Strong}} van Der {{Waals Type Rydberg-Rydberg Interaction}} in a {{Thermal Vapor}}},
  author = {Baluktsian, T. and Huber, B. and L{\"o}w, R. and Pfau, T.},
  year = {2013},
  journal = {Phys. Rev. Lett.},
  volume = {110},
  number = {12},
  pages = {123001},
  publisher = {American Physical Society},
  doi = {10.1103/PhysRevLett.110.123001},
  urldate = {2024-12-02}
}

@article{su2025,
  title = {Two-Photon {{Rydberg EIT}} Resonances in Non-Collinear Beam Configurations},
  author = {Su, Kevin and Behary, Rob and Aubin, Seth and Mikhailov, Eugeniy E. and Novikova, Irina},
  year = 2025,
  journal = { J. Opt. Soc. Am. B},
  volume = {42},
  number = {4},
  pages = {757},
  issn = {0740-3224, 1520-8540},
  doi = {10.1364/JOSAB.550937},
  urldate = {2026-03-12},
  langid = {english}
}

@article{su2022a,
  title = {Optimizing the {{Rydberg EIT}} Spectrum in a Thermal Vapor},
  author = {Su, Hsuan-Jui and Liou, Jia-You and Lin, I-Chun and Chen, Yi-Hsin},
  year = 2022,
  journal = {Opt. Express},
  volume = {30},
  number = {2},
  pages = {1499},
  issn = {1094-4087},
  doi = {10.1364/OE.444894},
  urldate = {2023-04-07},
  langid = {english}
}

@article{bai2020,
  title = {Self-{{Induced Transparency}} in {{Warm}} and {{Strongly Interacting Rydberg Gases}}},
  author = {Bai, Zhengyang and Adams, Charles S. and Huang, Guoxiang and Li, Weibin},
  year = {2020},
  journal = {Phys. Rev. Lett.},
  volume = {125},
  number = {26},
  pages = {263605},
  issn = {0031-9007, 1079-7114},
  doi = {10.1103/PhysRevLett.125.263605},
  urldate = {2023-04-07},
  langid = {english}
}

@article{Mohapatra2007,
  title = {Coherent Optical Detection of Highly Excited Rydberg States Using Electromagnetically Induced Transparency},
  author = {Mohapatra, A. K. and Jackson, T. R. and Adams, C. S.},
  journal = {Phys. Rev. Lett.},
  volume = {98},
  issue = {11},
  pages = {113003},
  numpages = {4},
  year = {2007},
  month = {Mar},
  publisher = {American Physical Society},
  doi = {10.1103/PhysRevLett.98.113003},
  url = {https://link.aps.org/doi/10.1103/PhysRevLett.98.113003}
}

@article{Tanasittikosol2011,
  author = {Tanasittikosol, M. and Pritchard, J. D. and Maxwell, D. and Gauguet, A. and Weatherill, K. J. and Potvliege, R. M. and Adams, C. S.},
  doi = {10.1088/0953-4075/44/18/184020},
  issn = {09534075},
  journal = {J. Phys. B: At. Mol. Opt. Phys.},
  number = {18},
  pages = {184020},
  title = {{Microwave dressing of Rydberg dark states}},
  volume = {44},
  year = {2011}
}

@article{schmidt2025,
  title = {All-{{Optical Radio-Frequency Phase Detection}} for {{Rydberg Atom Sensors Using Oscillatory Dynamics}}},
  author = {Schmidt, Matthias and Bohaichuk, Stephanie M. and Venu, Vijin and Wang, Ruoxi and K{\"u}bler, Harald and Shaffer, James P.},
  year = {2025},
  journal = {Phys. Rev. Lett.},
  volume = {135},
  number = {9},
  pages = {093602},
  issn = {0031-9007, 1079-7114},
  doi = {10.1103/23kb-7h7q},
  urldate = {2025-09-10},
  langid = {english}
}

@article{downes2020b,
  title = {Full-{{Field Terahertz Imaging}} at {{Kilohertz Frame Rates Using Atomic Vapor}}},
  author = {Downes, Lucy A. and MacKellar, Andrew R. and Whiting, Daniel J. and Bourgenot, Cyril and Adams, Charles S. and Weatherill, Kevin J.},
  year = {2020},
  journal = {Phys. Rev. X},
  volume = {10},
  number = {1},
  pages = {011027},
  issn = {2160-3308},
  doi = {10.1103/PhysRevX.10.011027},
  urldate = {2023-04-07},
  langid = {english}
}

@article{chen2022,
  title = {Terahertz Electrometry via Infrared Spectroscopy of Atomic Vapor},
  author = {Chen, Shuying and Reed, Dominic J. and MacKellar, Andrew R. and Downes, Lucy A. and Almuhawish, Nourah F. A. and Jamieson, Matthew J. and Adams, Charles S. and Weatherill, Kevin J.},
  year = {2022},
  journal = {Optica},
  volume = {9},
  number = {5},
  pages = {485},
  issn = {2334-2536},
  doi = {10.1364/OPTICA.456761},
  urldate = {2023-06-08},
  langid = {english}
}

@article{li2023,
  title = {Super Low-Frequency Electric Field Measurement Based on {{Rydberg}} Atoms},
  author = {Li, Ling and Jiao, Yuechun and Hu, Jinlian and Li, Huaqiang and Shi, Meng and Zhao, Jianming and Jia, Suotang},
  year = {2023},
  journal = {Opt. Express},
  volume = {31},
  number = {18},
  pages = {29228},
  issn = {1094-4087},
  doi = {10.1364/OE.499244},
  urldate = {2023-08-17},
  langid = {english}
}

@article{Jing2020,
author = {Mingyong Jing and Ying Hu and Jie Ma and Hao Zhang and Linjie Zhang and Liantuan Xiao and Suotang Jia },
title = {Atomic superheterodyne receiver based on microwave-dressed Rydberg spectroscopy},
journal = {Nat. Phys.},
volume = {16},
issue = {9},
pages = {911-915},
year = {2020},
doi = {10.1038/s41567-020-0918-5},
URL = {https://doi.org/10.1038/s41567-020-0918-5},
}

@article{Prajapati2021,
	author = {Prajapati, Nikunjkumar and Robinson, Amy K. and Berweger, Samuel and Simons, Matthew T. and Artusio-Glimpse, Alexandra B. and Holloway, Christopher L.},
	title = {Enhancement of electromagnetically induced transparency based Rydberg-atom electrometry through population repumping},
	journal = {Appl. Phys. Lett.},
	volume = {119},
	number = {21},
	pages = {214001},
	DOI = {10.1063/5.0069195},
	year = {2021},
}

@article{Holloway:2022qwb,
  title = {Rydberg Atom-Based Field Sensing Enhancement Using a Split-Ring Resonator},
  author = {Holloway, Christopher L. and Prajapati, Nikunjkumar and Artusio-Glimpse, Alexandra B. and Berweger, Samuel and Kasahara, Yoshiaki and Alu, Andrea and Ziolkowski, Richard W.},
  year = {2022},
  journal = {Appl. Phys. Lett.},
  shortjournal = {Appl. Phys. Lett.},
  volume = {120},
  number = {20},
  eprinttype = {arXiv},
  eprintclass = {physics.atom-ph},
  pages = {204001},
  issn = {0003-6951, 1077-3118},
  doi = {10.1063/5.0088532},
  url = {http://arxiv.org/abs/2202.08954}
}

@article{zhou2025a,
  title = {High-Sensitivity Rydberg Atom-Based Field Sensing Enhancement Using Miniaturized Resonator},
  author = {Zhou, Aojie and Lin, Yi and Mao, Ruiqi and Yang, Kai and Ding, Zhenke and Wan, Weipeng and Fu, Yunqi},
  year = {2025},
  journal = {IEEE Trans. Antennas Propag.},
  shortjournal = {IEEE Trans. Antennas Propag.},
  volume = {73},
  number = {12},
  pages = {10948-10952},
  issn = {0018-926X, 1558-2221},
  doi = {10.1109/TAP.2025.3596373},
  url = {https://ieeexplore.ieee.org/document/11123636/}
}

@article{deb2018b,
  title = {Radio-over-Fiber Using an Optical Antenna Based on {{Rydberg}} States of Atoms},
  author = {Deb, A. B. and Kj{\ae}rgaard, N.},
  year = {2018},
  journal = {Appl. Phys. Lett.},
  volume = {112},
  number = {21},
  pages = {211106},
  issn = {0003-6951, 1077-3118},
  doi = {10.1063/1.5031033},
  urldate = {2025-10-02},
  langid = {english}
}

@article{meyer2018,
  title = {Digital Communication with Rydberg Atoms and Amplitude-Modulated Microwave Fields},
  author = {Meyer, David H. and Cox, Kevin C. and Fatemi, Fredrik K. and Kunz, Paul D.},
  year = {2018},
  journal = {Appl. Phys. Lett.},
  volume = {112},
  number = {21},
  pages = {211108},
  issn = {0003-6951, 1077-3118},
  doi = {10.1063/1.5028357},
  url = {https://pubs.aip.org/apl/article/112/21/211108/35435/Digital-communication-with-Rydberg-atoms-and}
}

@article{hu2025,
  title = {Thermal Resonance-Enhanced Transparency in Room-Temperature Rydberg Gases},
  author = {Hu, Jinlian and Jiao, Yuechun and Yin, Yuwen and Lu, Cheng and Bai, Jingxu and Jia, Suotang and Li, Weibin and Bai, Zhengyang and Zhao, Jianming},
  year = {2025},
  journal = {Phys. Rev. A},
  shortjournal = {Phys. Rev. A},
  volume = {112},
  number = {2},
  pages = {L020801},
  issn = {2469-9926, 2469-9934},
  doi = {10.1103/yrs8-jg4r},
  url = {https://link.aps.org/doi/10.1103/yrs8-jg4r}
}

@article{zhang2026,
  title = {Coherent {{All-Optical Radio Frequency Phase Sensing Using Multiphoton Dressing}} and {{Interference}}},
  author = {Zhang, Hongqiao and Shen, Pinrui and Bohaichuk, Stephanie M. and Lippmann, Hanna and Kubler, Harald and Shaffer, James P.},
  year = {2026},
  journal = {arXiv preprint arXiv:2605.19851},
  url = {https://arxiv.org/abs/2605.19851},
  urldate = {2026-05-21},
  langid = {english},
  pubstate = {prepublished},
  version = {1}
}

@article{borowka2025,
  title = {Optically-Biased {{Rydberg}} Microwave Receiver Enabled by Hybrid Nonlinear Interferometry},
  author = {Borówka, Sebastian and Mazelanik, Mateusz and Wasilewski, Wojciech and Parniak, Michał},
  year = {2025},
  journal = {Nat. Commun.},
  shortjournal = {Nat Commun},
  volume = {16},
  number = {1},
  pages = {8975},
  issn = {2041-1723},
  doi = {10.1038/s41467-025-63951-9},
  url = {https://www.nature.com/articles/s41467-025-63951-9}
}
	
\end{document}